# Bayesian Uncertainty Quantification for Anaerobic Digestion models

Antoine Picard-Weibel[a,b,e,1], Gabriel Capson-Tojo[c], Benjamin Guedj[d,e], Roman Moscoviz[a]

[a]*SUEZ, CIRSEE,, 38 rue du President Wilson, 78230, Le Pecq, France*
[b]*Laboratoire Paul Painleve, Univ. de Lille, Cite Scientifique, F-59655, Villeneuve d'Ascq, France*
[c]*INRAE, Univ. Montpellier, LBE 102 avenue des etangs, 11100, Narbonne, France*
[d]*Centre for Artificial Intelligence, UCL 90 High Holborn, WC1V 6LJ, London, United Kingdom*
[e]*MODAL, Inria 40 avenue Halley, 59650, Villeneuve d'Ascq, France*

**Abstract**
Uncertainty quantification is critical for ensuring adequate predictive power of computational models used in biology. Focusing on two anaerobic digestion models, this article introduces a novel generalized Bayesian procedure, called VarBUQ, ensuring a correct tradeoff between flexibility and computational cost. A benchmark against three existing methods (Fisher's information, bootstrapping and Beale's criteria) was conducted using synthetic data. This Bayesian procedure offered a good compromise between fitting ability and confidence estimation, while the other methods proved to be repeatedly overconfident. The method's performances notably benefitted from inductive bias brought by the prior distribution, although it requires careful construction. This article advocates for more systematic consideration of uncertainty for anaerobic digestion models and showcases a new, computationally efficient Bayesian method. To facilitate future implementations, a Python package called 'aduq 'is made available.

## 1. Introduction

Computational modelling is now common practice in many areas of biology, including fields as diverse as environmental sciences, biotechnology and medical sciences (Sordo Vieira and Laubenbacher, 2022). These models are generally used to represent complex systems, such as gene regulation networks, ecological interactions, or biochemical reaction networks. In the field of environmental biotechnology, standard models based on ordinary differential equations have been adopted by the scientific community for the most common bioprocesses such as anaerobic digestion (AD) (Bernard et al., 2001; Batstone et al., 2002). Historically, modelers interested in AD have dedicated much effort over optimization techniques, in order to efficiently calibrate these models, notably gradient-based and derivative-free methods (see Donoso-Bravo et al. (2011) for a survey).

From a statistical viewpoint, the calibration of a number of parameters comparable to the number of observations is a high dimensional estimation problem, where special attention must be paid to the risk of overfitting. Rieger et al. (2012) indicate good practices for conducting validation, notably using different datasets, working under different conditions and operational parameters. In practice, observations can be scarce, preventing experimenters to fully validate their model (see Dochain and Vanrolleghem, 2001). This is notably the case when operating conditions are different for training and testing datasets. This assumption is called dataset or distribution shift in learning theory (Quinonero-Candela et al., 2008). In this setting, how much the predictions will diverge from the truth as time goes by cannot be inferred through validation, since validation data is assumed to be missing. Uncertainty quantification (UQ) methods are essential tools to assess the quality of the calibration. UQ is the quantitative analysis of the impact that sources of randomness have on the calibration process. This randomness originates from the measurements' noise as well as the stochastic nature of the future influent. UQ methods estimate how far the calibrated set of parameter values as well as the predictions of the calibrated model diverge from the truth, using statistical theory. Ideally, UQ on the predictions should be robust to distribution shift, warning the user whenever the calibration is no longer valid through an increase in predictions uncertainty.

---
[1] Corresponding author: antoine.picard.ext@suez.com

Unfortunately, careful assessment of uncertainty is far from systematic in AD modelling. Still, techniques have been used to quantify uncertainty for AD model calibration. The three prevalent methods in the field are based on Fischer's information matrix (FIM) (a.k.a. Cramér-Rao's lower bound or information inequality, see Chapter 2 in Lehmann and Casella, 1998), a statistical criterion introduced in Beale (1960) (from now on called Beale's method), and bootstrapping. These methods infer uncertainty on the parameters from residuals of the calibrated model. They are generally used to provide UQ only for parameters which are fitted, e.g., those selected by a sensitivity analysis routine. The remaining parameters are fixed at some default value, as they are not deemed to have sufficient influence on predictions. But those parameters actually have large uncertainty, since the data is not able to discriminate between two widely different values. As the sensitivity of the model's response to each parameter depends on operating conditions, the quality of extrapolations of the model on new conditions can only be known if uncertainty is quantified on all parameters, or if these operating conditions have been previously validated (Rieger et al., 2012).

Bayesian methods are designed to tackle the uncertainty on all parameters throughout the training process, as they perform calibration and UQ jointly. The uncertainty on parameters with little sensitivity impact on the model is controlled through expert's knowledge, encoded in a probability distribution called prior. The prior is twisted into the posterior distribution through confrontation with the observations, concentrating on sets of parameters likely to have generated them. The calibrated model is no longer deterministic, but stochastic.
While uncommon, Bayesian flavored techniques have already been used in the context of AD modeling (Martin et al., 2011; Couto et al., 2019; Pastor-Poquet et al., 2019). All these Bayesian inspired algorithms output the uncertainty in the form of a sample. Statistical theory shows that satisfactory description of a generic distribution requires a number of samples increasing at a more than exponential rate with the number of parameters fitted (Goldenshulger and Lepski, 2014). This heavily restricts the applicability of such algorithms for highly parameterized models.
Variational Bayesian methods (Hinton and Van Camp, 1993; Beal, 2003) learn the best approximation of the posterior amongst a class of distributions (e.g., Gaussians, Gaussians mixture). This structured posterior can be fully assessed using fewer model evaluations. The flexibility of the posterior (covariance structure, multimodality) is governed by the distribution class considered. For AD modelling, Gaussian posteriors with block diagonal covariance present an interesting tradeoff, exploring nontrivial correlation structure while limiting the number of hyperparameters.
A novel methodology for the joint calibration and UQ of AD models based on variational Bayes and learning theory, called Variational Bayesian Uncertainty Quantification (VarBUQ), is introduced. The methodology is available as an open-source Python package and can be readily applied to any AD or biochemical reaction network model. Its performance is compared to the prevailing ad hoc UQ routines (FIM, Beale and Bootstrap), considering two AD models of varying complexity - Anaerobic Model 2 (AM2, Bernard et al. (2001)) and Anaerobic Digestion Model 1 (ADM1, Batstone et al. (2002))-, using six synthetic datasets describing three different operating conditions and two sets of parameters monitoring the prior distribution's inductive bias. Such synthetic data allows to assess the ability of the UQ methods to recover the true parameter values, as well as their performance on test datasets. Special attention is paid to the robustness of each UQ methods to distribution shift.

## 2. Material and methods
*2.1. Anaerobic Digestion models and data generation*

Observations used for calibration are generated using the AD model considered in each benchmark (ADM1, AM2). The description of the intrant is likewise synthetic. ADM1's code is adapted from the PyADM1 package https://github.com/CaptainFerMag/PyADM1. AM2's implementation was written from the formulation given in Bernard et al. (2001), with slight modifications to benefit from pH measurements and include microbial mortality. Both models are implemented in Python and are accessible on the following github repository: https://github.com/BayesianUncertaintyQuantifAnaeroDig/AnaeroDigUQ.

For each AD model, the modelled digester consists in a single stage digestion system with a liquid phase measuring 3400 cubic meter and a gas phase measuring 300 cubic meters. The temperature inside the digester is kept constant at 308.15 K. The frequency of the data was set to represent realistic monitoring data - respectively 24 hour and 6 hours between each data point for observations (e.g., biogas quality) and influent data (e.g., mass flow). 280 days are simulated, with the first 70% (196 days) used for calibration. The remaining part is used to assess the validity of the UQ methods on the predictions. Each AD model is used to generate four datasets, describing two levels of intrant dynamism and two different sets of parameters. The datasets are denoted LN, HN, LF, HF, with L and H standing respectively for low and high dynamism of intrants, while N and F stands respectively for a set with parameter values near or far their default values. The same intrant was used for L datasets (respectively H datasets), while the same set of parameters is used to construct N datasets (resp. F datasets). The N and F parameters are constructed using the prior distribution (see Section 2.2.1). The N (respectively F) parameter is drawn from the prior, and then renormalized in such a way that the $\chi^2$-test rejects the hypothesis that it is drawn from the prior with a p-value of 0.95 (resp 0.05). The intrant datasets are constructed using sums of random sinusoids. The L and H datasets differ in terms of minimum HRT (30 days for L, 10 days for H). The intrant is assumed to contain only carbohydrates, proteins and lipids, the remaining concentrations being set to 0. The corresponding datasets are available in the github repository.

To assess the robustness of the calibration and UQ methods under change of operating conditions, two additional datasets for each AD model are constructed by modifying the HRT of the L datasets during the test period. The HRT was lowered to 10 days gradually over two weeks. These datasets with varying levels of intrant dynamism are denoted LVN and LVF. A specific initial state is constructed for each dataset by considering the steady state result of the AD model, using the first characterization of the intrant. After data has been generated, a noisy copy of inputs, observations and initial states are used for the calibration and UQ routines. The signals are noised in log-space using uniform noise:

$$\log(\text{Obs}_{i,t}) = \log(\text{AD}^*_{i,t}) + U(-\sigma, \sigma)$$

For both AD models, the influent's noise level is 0.08, while the observations noise level is 0.15. The noise levels and the distribution (log-uniform) are not assumed to be known during the optimization and UQ stages. Yet, the algorithms do rely on the assumption that the observations are noised in log-space. As such, the statistical model considered when performing FIM and Beale's UQ is

$$\log(\text{Obs}) = \log\bigl(\text{AD}(\theta^*, \text{Influent}, \text{Obs}_0)\bigr) + \sigma\epsilon$$

Considering the influent description, none of the UQ method assumes any noise. Yet, noise was still added to the influent data. Such noise is transformed in a nonlinear way by the AD model into noise on the predictions. This makes the overall noise structure more complex and closer to real world scenario - while the statistical model considers a simplified, more tractable noise structure.

Consistently with this model, the objective function or score function considered is essentially a root mean square error in the log space, with slight adjustment to improve stability (see documentation of github repository for details). The score can be roughly interpreted as the mean relative error amongst all the predictions for small score values, i.e. a score of $S \ll 1$ can be thought of as relative mean error of $100 \times S\%$.

Predictions considered to compute the score for AM2 are the concentrations of soluble compounds ($S_1$, $S_2$ in the original paper) and the gas flows ($q_M$ and $q_C$) (784 observations), while the predictions used for ADM1 were VFA concentrations ($S_{va}$, $S_{bu}$, $S_{pro}$, $S_{ac}$), concentration of inorganic nitrogen ($S_{IN}$), gas flows ($q_{gas}$, $q_{CH4}$) and partial pressures ($p_{gas,CH4}$, $p_{gas,CO2}$) (1764 observations).

*2.2. VarBUQ algorithm*

2.2.1. Construction of the prior

Being a generalized Bayesian algorithm, VarBUQ requires a description of the a priori belief on the values of parameters in the form of a probability distribution, known as the prior distribution. For both AD models, the prior distribution consists in multivariate Gaussian distributions with a diagonal covariance structure on the log parameters.

As such, the global prior draws each parameter independently from one another. This follows the recommendations of Tsigkinopoulou et al. (2017). The individual priors are constructed using previous description of default values and uncertainty (Rosén and Jeppsson (2006); Batstone et al. (2002) for ADM1, Bernard et al. (2001) for AM2). Means and standard deviations of the priors can be found in the supplementary material, while complete methodological details are given in the documentation of the github repository.

2.2.2. Variational formulation of generalized Bayes

From a prior distribution, the Bayesian framework constructs a posterior distribution by taking into account the observed data. Formally, it requires a statistical model, in the form of a likelihood function $\ell(\theta, Obs)$, which measures the affinity between the set of parameters and observations. The posterior is then defined using Bayes' formula, as

$$\hat{\pi} = \pi(\theta \mid \text{Obs}) \propto \ell(\theta, \text{Obs})\pi_0(\theta)$$

A celebrated result of the Bayesian framework is that, when the number of parameters is finite, the posterior distribution's credible regions asymptotically behave like true confidence regions, under mild assumptions on the model and the prior (Bernstein-von Mises's theorem, see chapter 12 of Le Cam, 2012). This implies that the posterior concentrates around the true set of parameters, and that high probability regions for the posterior can be used to adequately quantify the uncertainty on the parameters, with asymptotical guarantees that these regions will have the required coverage. Unfortunately, the Bayesian framework is not robust to improper statistical modelling, as is the case in the context of AD. To address this shortcoming, different variants, such as tempered posteriors and Probably Approximately Correct (PAC)-Bayes, were introduced (see Guedj, 2019, for a survey). A generic methodology popular in environmental modelling, the Generalized Likelihood Uncertainty Estimation (GLUE) framework, advocates replacing the modelling-based likelihood function by some user chosen proxy (Beven, 2018). This core idea is also shared by the more theory-oriented PAC-Bayesian paradigm, where the likelihood is replaced by the empirical prediction error, which acts as the standard measurement of affinity between data and set of parameters in learning theory. A thoroughly studied case is the Gibbs posterior, defined through

$$\hat{\pi}_\lambda(\theta) \propto \exp\left(-\frac{S(\theta)}{\lambda}\right)\pi_0(\theta)$$

By analogy with thermodynamics, $\lambda$ is called the PAC-Bayesian temperature. It should not be confused with the temperature in the AD process, to which it is not related. It controls the amount of trust given to the observations: for large PAC-Bayesian temperatures, the posterior is close to the prior, while for low PAC-Bayesian temperatures it is similar to a Dirac distribution putting all its probability mass on the minimizer of the score – the celebrated empirical risk minimizer.

While Bernstein-Von Mises's theorem is no longer valid, some guarantees do exist for PAC-Bayesian posteriors. These comes in the form of probably approximately correct (PAC) generalization bounds, controlling the mean error for the posterior on unseen data. A variety of such bounds were constructed, first for independent, identically distributed observations with bounded errors (McAllester, 1999; Seeger, 2002; Catoni, 2007), then under less stringent assumptions (Alquier and Wintenberger, 2012; Seldin et al., 2012; Alquier and Guedj, 2018; Haddouche and Guedj, 2022). However, the lack of an accepted model for intrant distributions makes it difficult to assess the validity of the assumptions even of the most general PAC-Bayes bounds. As such, the PAC-Bayesian technique used does not come with any theoretical guarantee. Rather than relying on Markov chain Monte Carlo techniques which would construct samples from the posterior, VarBUQ is inspired by the variational inference framework, computing hyperparameters which define a distribution. The algorithm is based on Catoni's PAC-Bayes bound (Catoni, 2007), which interprets Gibbs posteriors as result of penalized optimization:

$$\hat{\pi}_\lambda = \arg\inf_{\nu \ll \pi} \nu[S] + \lambda \mathrm{KL}(\nu, \pi_0)$$

where KL denotes the Kullback–Leibler divergence. This penalization results in the output (the posterior distribution) diverging from the input (the prior distribution) only if the data shows that this is necessary. In other words, there is a tradeoff between "fitting to the observations" and "remaining close to the prior." As such, the prior gives inductive bias: it tells the algorithm where it should look for solutions, and how likely one deems each potential solution. This is different to the standard empirical risk minimization algorithm, which does not come with inductive bias: the algorithm will pick up any set of parameters, regardless of whether it would be deemed plausible or not by an expert.

To obtain a result that is interpretable, computable and that can be easily saved and reused, the minimization problem is reduced to a parametric family of distributions. Gaussian distributions with a covariance matrix satisfying some assumptions are considered. These distributions are defined on the unconstrained parametrization - as the transform between this representation and the standard parametrization is bijective, there is no difficulty in interpreting these distributions as distributions on the standard set of parameters. In order to keep the dimension of the parametric family of Gaussian reasonable, the covariance matrices were constrained to be block diagonal. The blocks are constructed by clustering parameters having direct impact on the same reactions (see documentation of the github repository). Parameters belonging to different blocks are drawn independently from one another. The Bayesian calibration problem is therefore simplified to

$$\hat{\pi}_\lambda = \arg\inf_\gamma \pi(\gamma)[S] + \lambda \text{KL}(\pi(\gamma), \pi_0)$$

where γ is the hyperparameter describing the distribution. This is solved using accelerated gradient descent. Indeed, since the prior is also Gaussian, the KL admits a closed form expression, whose gradient with respect to the distribution's parameters can be computed explicitly. The derivative of the integral with respect to π(γ) can be estimated using an independent sample of sets of parameters θ drawn from π(γ).

This estimate is unbiased, and its variance scales in the inverse of the number of samples points. The number of calls to the AD model, which is usually the computational bottleneck (Rosén and Jeppsson, 2006), equals at each step the number of sample points generated. These evaluations can be fully parallelized. To be able to keep this number of model evaluation reasonable, mechanisms are used to recycle the evaluations (see documentation of the github repository). For each dataset, the gradient descent procedure runs for 250 steps. For AM2 (resp. ADM1), 256 (resp. 160) samples are generated at each step, amounting to 64000 (resp. 40000) calls to the AD model. A full description of the hyperparameters is given in the documentation of the github repository. The choice of the PAC-Bayesian temperature is done a priori. The criteria used to define this PAC-Bayesian temperature is based on Catoni's PAC-Bayes generalization bound (Catoni, 2007). This bound, valid for scores defined as a mean of $N$ independent, identically distributed losses bounded by $C$, states that

$$\mathbb{E}_S[\nu[S]] \leq \nu[S] + \lambda KL(\nu, \pi_0) + \frac{C}{\lambda 8N} + \lambda \log \delta^{-1}$$

is valid simultaneously for any distribution ν with probability at least $1 - \delta$. The generalization guarantee involves the term $\frac{C}{\lambda 8N}$ This implies vacuous generalisation guarantees if the PAC-Bayesian temperature chosen is too low. The PAC-Bayesian temperature is chosen in such a way that $\frac{1}{\lambda 8N} < 0.1$, which implies for ADM1 that $\lambda \geq 0.0007$ and for AM2 that $\lambda \geq 0.0016$. A safety margin was added, and PAC-Bayesian temperatures of $\lambda_{\text{ADM1}} = 0.001$ and $\lambda_{\text{AM2}} = 0.002$ were used. For ADM1, PAC-Bayesian temperatures two and eight times larger were also investigated. It should be stressed that this a priori choice of PAC-Bayesian temperature is debatable, since Catoni's bound assumptions are not met: data is not independent, not identically distributed, and the score was not a sum of contributions, but the square root of a sum.

For ADM1, the initial distribution's parameters were obtained through a specific algorithm, which was able to efficiently reduce the mean score. This was based on the approach described in Leurent and Moscoviz (2022), which uses large samples of sets of parameters to completely redefine Gaussian distributions. The procedure stops when the objective in Equation (4) starts increasing. Such an approach proved necessary since the gradient descent procedure struggled for ADM1 to quickly concentrate the distribution around a satisfactory set of parameters when initiating from the prior.

*2.3. Other uncertainty quantification routines included in the benchmark*
Three UQ routines are considered to benchmark VarBUQ: FIM, Beale and Bootstrap. For Bootstrap, the implementation was based on Regueira et al. 2021. A comparison of the three routines is provided in the supplementary material. Bootstrap was not evaluated for ADM1, due to excessive computation time.

Contrary to Bayesian joint UQ and calibration, these UQ routines are carried out after model calibration. This non-Bayesian calibration was performed by minimizing the score function (using CMA-ES, Igel et al. 2007). As these methods do not follow the Bayesian paradigm, they do not require constructing a prior distribution and therefore can be easier to implement.

Selecting parameters to calibrate through sensitivity analysis can mitigate the risk of overfitting. The more parameters are selected, the smaller the empirical score of the calibrated model will be. For ADM1, a global sensitivity analysis based on Morris method (Morris, 1991) is performed to select the parameters to calibrate (with 96 repetitions). The minimum and maximum values considered for each parameter are coherent with the prior used by VarBUQ, being two standard deviations below and above the reference value. Parameters whose sensitivity values were above 0.025 were selected for calibration. Details on the implementation can be found in the documentation of the github repository.

*2.4. Assessment of parameters uncertainty*

AD model parameters describe quantities which have a physical or biological interpretation and inform on properties of the AD process. As such, the uncertainty on the calibrated parameter values is an important consideration.

Each UQ method is assessed through computation of p-values for tests of the hypothesis that the true set of parameters is $\theta^*$. If the UQ method performs as it should, these p-values should be uniformly distributed between 0 and 1. Small p-values indicate that the UQ is overconfident, as the true set of parameters would be rejected, while large p-values indicate that the UQ is underconfident.

*2.5. Assessment of uncertainty on predictions*

The performance of each UQ method is also assessed on predictions using the test set (84 days). The uncertainty on the prediction is obtained by transferring the uncertainty on the parameter, through linear uncertainty transport for FIM, and through the evaluation of multiple sets of parameters for all remaining methods. Pseudo 95% confidence intervals (CIs) were then constructed for each prediction by considering quantiles, and their ability to cover the unnoised signal is assessed. Predictions are regrouped as gas flows ($q_M$, $q_C$ for AM2, $q_{gas}$, $q_{CH4}$ for ADM1) and soluble compounds ($S_1$, $S_2$ for AM2, the four main VFAs, $S_{bu}$, $S_{va}$, $S_{ac}$, $S_{pro}$ for ADM1) to assess quality.

For each group of predictions, four indicators are computed:

- The coverage of the CIs, i.e. the fraction of predictions inside the CIs,
- The width of the CIs,
- The prediction error of the calibrated model,
- The residual error of the CIs.

The residual error of the CI is computed by replacing the standard residuals by the distance between the ground truth and the CI. Notably, if the confidence interval completely covered the truth, the residual error of the CI would be 0.

## 3. Results and discussion

*3.1. Calibration results on training set*

For ADM1, the global sensitivity analysis selected from 9 to 14 parameters depending on the datasets. Only 14 parameters were at least selected once ($K_{S\_c4+}$, $k_{m\_c4+}$, $K_{S\_ac}$, $k_{m\_ac}$, $K_{S\_pro}$, $k_{m\_pro}$, $k_{m\_aa}$, $k_{m\_su}$, $k_{dec}$, $K_{I,NH3}$, $pH_{UL:LL\ aa}$, $pH_{LL\ aa}$, $pH_{UL:LL\ ac}$, $pH_{LL\ ac}$ in the original paper). Details on parameters selected for calibration for each dataset can be found in the supplementary material.

Once calibrated, the models obtained scores of about 0.09 for AM2 and 0.095 for ADM1. This is slightly above the contribution of the noise on the observations (theoretically, 0.087 based on the selected noise level on observations alone). This implies that the noise on the influent did increase the overall noise on observations. As expected, the optimization-based calibration routine succeeded in finding sets of parameters achieving a lower score than the one obtained using the true set of parameters. The mean score for VarBUQ is slightly above the score of the true set of parameters for all datasets at the reference PAC-Bayesian temperature (about 0.005 higher for both AM2 and ADM1, implying an absolute increase of 0.5% to the relative error). Doubling the PAC-Bayesian temperature had moderate effect on the train performance of the posterior, with an increase in the score of about 0.005. Increasing the PAC-Bayesian temperature to eight time its reference value had a more noticeable effect, with a mean score up to 0.024 higher.

## 3.2. Uncertainty on parameters values

The capacity of the UQ methods to capture the true set of parameters was assessed by computing p-values for tests indicating whether the true set of parameters belonged to the confidence regions. These p-values are tabulated in Table 1.

Ad-hoc confidence regions constructed after standard calibration could generally not account for the large deviations between the true set of parameters and optimized set of parameters for ADM1. This results in FIM's confidence regions systematically failing to cover the true set of parameters for ADM1, where deviations are particularly noticeable for $k_m$, $K_S$ couples. This finding remains mostly valid in the case of AM2, since the confidence level must be chosen above the standard 95% criteria in order to cover the true set of parameters with FIM and Bootstrap confidence regions. The results of Beale's method are of particular interest. As the p-values were constructed using the theoretical criteria rather than any approximation, its failure to englobe the true set of parameters directly implies that the non-linearity in the AD models offers opportunities to reduce the noise significantly more than a linear model. Half of the p-values obtained were orders of magnitude lower than the 0.05 threshold considered, being on two occasions equal to the machine precision ($2.2e-16$). On the remaining datasets, only one p-value was above the threshold (0.3), while the three others were of order $1e-3$. Confidence regions constructed with Bootstrap failed to cover the true parameter for any confidence level in three of four cases. Correcting for the number of bootstrap samples generated, the 95% confidence upper bound on the p-values was above the standard 0.05 threshold for only one dataset out of four. This could be related to specific implementation choices designed to mitigate the computation time (see documentation of the github repository). Bootstrap methods are by construction computationally intensive, requiring multiple model calibrations, which in the context of AD might be prohibitive. The computational cost of the bootstrap routine could be improved by considering different calibration techniques or laxer termination criteria. However, no satisfactory tradeoff between performance and computational cost was found during the present study.

Of all UQ methods, VarBUQ gave the best results for parameter recovery. For ADM1, the results remained unsatisfactory. Using the reference PAC-Bayesian temperature, only one p-value was above the threshold (compared to none for all remaining methods), while two of them were equal to the machine epsilon, and the last one of order 1e-6. This was improved by doubling the PAC-Bayesian temperature, which brought little train performance loss, though there was still only a single p-value above the threshold. p-values were further increased by raising the PAC-Bayesian temperature to eight times the reference, at the cost of noticeable decrease on train performances. In that last setting, two p-values were above the threshold, while the two remaining ones are of order 4e-3. For AM2, VarBUQ with reference PAC-Bayesian temperature obtained satisfactory performance, with p-values all of order 0.9, both for L and F datasets. For the latter ones, the prior would obtain p-value of 0.05, implying that the posterior did more than inherit the induction bias.

Plots of the confidence regions constructed through each UQ method for the AM2 model (Figure 2) yield qualitative insight on their performances. VarBUQ benefits from inductive bias as exhibited in figs. 2a, 2b and 2e to 2h, where the confidence region constructed using the posterior remains almost entirely in the confidence region constructed using the prior. Still, VarBUQ performed satisfactorily in settings where the true set of parameters is on the boundary of the prior's confidence regions (figs. 2c and 2g). The remaining UQ methods are almost ordered, with FIM's confidence region nearly englobing Beale's, which in turn englobes those constructed by the Bootstrap method. While this seems incoherent with the p-values obtained in table 1, this could be explained by the additional approximation step required to construct Beale's confidence regions. While all UQ method indicate strong correlation between maximum growth rate and Monod constant, the exact form of the confidence regions differs. By construction, FIM's confidence regions are ellipsoidal, while VarBUQ's confidence regions are ellipsoidal in log-space. Interestingly, this second shape-constraint seems better suited to describe the relationship between the two parameters, since both Beale and Bootstrap, which outputs confidence regions with no shape constraints, obtain a somewhat similar curvature (figs. 2b to 2f and 2h). Since FIM's confidence regions are constructed by extrapolating a local linear approximation, they can include nonphysical parameter values (i.e. negative values), as is the case in figs. 2e and 2f and, to a lesser degree, in fig. 2g.

Overall, Figure 2 highlights two factors which contribute to VarBUQ's superior performance in comparison to the ad-hoc UQ methods for the AM2 datasets. First, the confidence region it constructed tend to be larger than those constructed using other UQ methods. Second, VarBUQ's confidence regions are also better centered around the true parameter values, implying that the Bayesian procedure offered a better calibration than the standard calibration procedure. This second feature can be attributed to the inductive bias brought by the prior: out of two sets of parameters yielding similar outputs, VarBUQ will favor the one deemed most likely by the experts (i.e., encoded in the prior), even if slightly less performant.

*3.3. Uncertainty on prediction values*

As complex AD models such as ADM1 are known to have identifiability issues, assessing the performance of the UQ on the parameter is not sufficient. Indeed, since different sets of parameters may still result in similar predictions, confidence regions centered around an incorrect set of parameters could still encapsulate the uncertainty on the predictions. Still, recovering the true set of parameters is the only way to provide full guarantees on the performance of the model on any future dataset.

Amongst all UQ methods tested, VarBUQ was best able to recapture the underlying signal. The coverage of its 95% confidence intervals is significantly higher than the other methods for both AD model (see Figure 3b), achieving an overall mean coverage of 69%, compared to 38% for FIM, 35% for Beale and 30% for Bootstrap. For each method, the CIs obtained higher coverage for soluble compounds than gas flows - this can be explained by the smaller sensitivity of gas flows to parameter values, as non-biological gas-related parameters such as $K_p$ were assumed to be constant, and higher sensitivity to the input noise. The higher coverage obtained by VarBUQ is coherent with the larger width of its CIs (see Figure 3d). While for the gas flows, these remain too small to fully capture the signal at the target 95% level, VarBUQ's CIs can be large for soluble compounds (up to $\pm 10\%$ for AM2 and $\pm 17.5\%$ for ADM1). Notably, for AM2, VarBUQ's CIs systematically had 100% coverage, indicating underconfidence in the results. Other UQ methods constructed CIs achieving above 95% coverage which were more than 3 times smaller for one dataset. Oracle symmetric confidence intervals achieving 100% coverage could be up to 5.6 times smaller, implying that the methodology can still be improved upon. Still, no other UQ method was able to obtain consistently high coverage for soluble compounds. For ADM1, the large width of VarBUQ's CIs appears necessary to obtain the required coverage. In the single case where another UQ method covered more than 90% of the data (Beale for acetate concentration, HN), CIs' width was larger than the one obtained by VarBUQ (0.29 vs. 0.23), for lower coverage (92% vs. 95%).

VarBUQ was able to maintain a high level of coverage when the operating conditions change, for a reasonable increase in the width of the CIs. CIs constructed using Fisher's information matrix reacted drastically to changes of operating conditions. The width of the CIs reached very high levels for S2 in the AM2 model (resp. 2.2 and 3.0 for LVN and LVF, implying an average factor of 8 and 20 between the lower and upper bound on the prediction). This phenomenon is also observed, though to a lesser degree, for ADM1. This could be explained by FIM using linear extrapolation of local changes in the predictions, which do not take into account saturation effects. This interpretation is corroborated by the fact that the widths of Beale's CIs do not evolve in such a way - resulting in a drop in coverage.

Both the standard calibration and VarBUQ obtained low errors on test sets similar to the train sets, with average prediction errors remaining lower than 0.021 for AM2, 0.04 for ADM1. While VarBUQ slightly underperformed when the test sets were similar to the training sets (obtaining prediction scores on average 15% higher), it exhibited stronger robustness to change of operating conditions. For those datasets, the average prediction errors of the standard optimization were 0.041 for AM2 and 0.097 for ADM1, respectively 53% and 84% higher than the prediction errors of VarBUQ. The residual prediction error computed after projecting on the CIs is globally smaller using VarBUQ (see Figure 3c). Notably, it was the only method able to obtain residual prediction errors to a low level ($< 0.05$) for all predictions. Since the predictions are already small when the test influent is similar to the train influent, this indicator is more relevant for the datasets with distribution shift. For AM2, Bootstrap, Beale and FIM obtained their worst performance on the same variable, S1 (for LVN dataset), of respectively 0.15, 0.10 and 0.087, indicating a sizeable gap between the signal and the CIs. For ADM1, Beale and FIM obtained non negligible residual prediction error for the concentrations of acetate (0.22 and 0.18 respectively). Beale's UQ also failed to properly account for the propionate (0.19).

Overall, VarBUQ's UQ was best able to capture the discrepancy between the predictions and the true signal. Both FIM and Beale slightly outperformed the Bootstrap method. While FIM and Beale obtained similar performances, FIM reacted better to the change in intrant characteristic, obtaining higher level of coverage.

Figure 4 represents the CIs for the predictions of the three main VFAs (butyrate, propionate and acetate) obtained by VarBUQ, FIM and Beale on ADM1, for LN and LVN datasets. Without distribution shift (figs. 4a to 4c), the CIs constructed by VarBUQ englobe those constructed through FIM, which were on average larger than those constructed by Beale's method. All CIs exhibited high frequencies, due to the noisy intrant description. While only VarBUQ's CIs adequately covered the true signal, the remaining methods still obtained satisfactory performances as the calibrated model's predictions were suitable. Under distribution shift (figs. 4d to 4f), the calibrated model's prediction diverged significantly from the truth. FIM's CIs widened sufficiently to take into account this discrepancy for butyrate and propionate concentrations, but not for acetate concentrations. The width of Beale's CIs remained noticeably too small. On the other hand, VarBUQ's CIs were centered around the true signal, and englobed it adequately.

*3.4. Computational cost*

Computations were carried out using Microsoft Azure, on virtual machines with 32 cores, 64 Gb ram and 256 Gb of memory. Routines fully benefit from parallelization and one can assume that the number of cores have an almost linear impact on their durations. All durations are supplied in supplementary materials.

VarBUQ was more computationally intensive than the standard calibration, requiring an average of 1 h 40 minutes for AM2 (resp. 30 minutes for standard calibration) and 5 hours for ADM1 (resp. 2 h 30 minutes). This is mitigated once UQ is taken into account. While FIM method's duration is negligible, Beale's method required 50 minutes for AM2 and more than an hour and a half for ADM1, bridging a large part of the gap. The Bootstrap procedure required prohibitive computational power. As such, this method was only assessed for AM2, with computations lasting about two days. While this computation time could be diminished, by either by reducing the number of Bootstrap procedure or relaxing convergence criteria, this would have serious consequences on the quality of the UQ.

*3.5. Potential bias related to calibration method*

The performance of the UQ methods benchmarked are impacted by the calibration method. As such, the empirical risk minimization approach used here should be deemed in part responsible for the obtained results. This choice of calibration method was driven by two considerations. First and foremost, it is a quite common approach in the field, and therefore the results are hopefully representative of the difficulties of obtaining proper UQ for AD models. A second point is that Beale's UQ method takes its origin in the behavior of the minimizer of mean squared errors objective. To limit confounding factors when assessing the UQ methods, the calibration derived from Beale's method was therefore used also for FIM and Bootstrap, while VarBUQ uses the same scoring function.

For FIM, such a calibration is actually ill-suited to the method's hypothesis, since the requirement that the estimator be unbiased is not met. However, it should be stressed that this hypothesis will rarely be realistic in the context of AD models, most of all for highly parametrized models. Constructing an unbiased estimator might not be feasible, even when considering a simple statistical model such as Equation (2) - and since the statistical models used have only limited validity, there is little point in trying. Moreover, from a statistical viewpoint, the well-known bias–fluctuation tradeoff indicates that biased estimator can give better performances.

One important difficulty with the optimization-based procedure used was that it could result in unrealistically high parameter values for the $k_m$, $K_S$ couples. This was treated by imposing an upper bound on those values when optimizing. This could explain why the optimization procedure had poor robustness when testing on a different intrant. This hints that the calibration could benefit from penalization, in order to favor explanations remaining closer to the standard values.

*3.6. Limitations of synthetic datasets*

Knowledge of the true set of parameters being primordial when assessing UQ methods for parameter recovery, the benchmark was conducted using synthetic datasets. This implies some debatable modelling decisions. A first decision concerned the modelling of the noise. The signal was noised in log space. While not strictly accurate, this implies that the measurement noise will typically be better represented considering relative error. A strong hypothesis was to use the same noise level for all types of observations. This is actually a key requirement in order to use Beale's UQ technique when using different types of predictions. Adapting Beale's method when the noise levels vary is not straightforward, as a core aspect of the method consists in bypassing the estimation of the noise levels. A uniform noise structure was preferred to the standard Gaussian noise, to test whether this slight change would give an edge to the Bootstrap procedure, specifically designed to deal with unknown noise structure. Noise on the input data resulted in prediction CIs with little smoothness. While the influent signals could have been smoothed, this could have added less detectable biases to the analysis of the results (e.g., choice of the smoothing bandwidth). In practice, observations on the influent might be scarce or exhibit high frequency noise, and as such, the modelling did not seem too unrealistic.

Another consideration is that the performance of calibration and UQ using real world data will depend on the mismatch between the computational model and the physical model. Still, the experiments conducted inform on the quality of the UQ methods. A method struggling with synthetic data is unlikely to fare better with real world data. Finally, the methodology used to construct the true set of parameters might favor the Bayesian framework, insofar as the prior is used. This was mitigated by assessing the performance on a set of parameters which was deemed unlikely to have been drawn from the prior (p-value of 0.05). Still, any Bayesian framework is expected to work poorly if the prior is badly constructed and is either much too large (resulting in underconfidence) or much too small (resulting in both poor calibration performances and overconfidence). Constructing adequate priors is therefore a key challenge for the use of Bayesian methods with real data. Thorough bibliographical work is needed to make use of numerous previous works and obtain a state of the art prior. The benchmark's results show that such work could prove valuable; although the other UQ methods can be implemented more easily as they do not require a prior, the Bayesian procedure benefitted from the prior, obtaining confidence regions better centered around the true sets of parameters and confidence intervals on predictions more robust to distribution shift. Notably, it prevents including sets of parameters which an expert would consider unrealistic.

*3.7. VarBUQ compared to previous Bayesian routines for Anaerobic Digestion*

Before the present work, Bayesian flavored techniques had already been used in the context of AD modelling. Martin and Ayesa (2010) developed a Matlab implementation of Monte Carlo methods which could calibrate a 2-parameters AD model accurately while also assessing parameter uncertainty, adapt to non-identifiable situations, as well as construct proper and tight confidence regions for predictions (Martin et al., 2011). Couto et al. (2019) use a Bayesian framework to fit five parameters in ADM1. Pastor-Poquet et al. (2019) implemented an ad hoc Approximate Bayesian Computation (ABC) algorithm to calibrate 14 parameters on a high-solids AD model. Due to implementation choices, the actual algorithm's UQ presents characteristics between Beale and Bayesian methods. The resulting mean parameter was found to offer good predictive power for methane production, though the authors also noted discrepancies in volatile fatty acids (VFA) simulations which could be due to modelling issues.

These Bayesian inspired algorithms output the uncertainty in the form of a sample. Conversely, VarBUQ does not output a sample. The algorithm computes hyperparameters defining a probability distribution belonging to a user chosen parametric class (e.g., multivariate Gaussian). This offers a more interpretable description of the uncertainty, able to effortlessly generate any number of samples from the posterior. This description can be furthermore stored and used for further calibration, assuming more data has been collected; as such, the algorithm can easily be used in an online learning set-up. In addition, VarBUQ considers a simplified Bayesian framework limiting the interactions between parameters to specific cases, chosen through expert knowledge. For instance, in this study, only interactions between parameters acting on the same biological reaction were allowed. This was implemented by considering gaussian distributions with block diagonal covariance, which significantly reduce the number of hyperparameters (e.g., 54 versus 465 for ADM1). This more rigid set-up limits the ability of the posterior to fit the data but reduces the number of model evaluations needed compared to learning a full covariance matrix or general distribution.

*3.8. Improving VarBUQ*

The Bayesian paradigm showcased here can be improved both in terms of methodology and implementation. A key aspect is the construction of the prior, which could take into account observed correlations between parameters. This would help the posterior further concentrate by removing unlikely combinations of parameters. Another leverage for improvement is the procedure choosing the PAC-Bayesian temperature. Informally, the choice of PAC-Bayesian temperature should be guided by how much training data is used and how far the data ought to be trusted. Quantifying this confidence would be much harder in real world scenario. Selecting the PAC-Bayesian temperature through validation could be a computationally costly option.

The computational cost of the procedure could be reduced. A promising option consists in building surrogate models able to approximate the error of the model for a fraction of the computational cost. Increasing the maximal time step of the ordinary differential equation solvers could be a simple option. This could help quickly building good approximations of the posterior.

The variational class plays an important role both in terms of computational complexity and performance. The choice investigated here, Gaussian distributions with block diagonal covariance matrix, appeared a good compromise. The block covariance structure prevented the posterior from learning spurious correlations between variables, while it was still able to investigate non identifiable cases. Gaussian distributions are also easier to manipulate compared to the prevalent choice in the AD literature, where a combination of uniform and log-uniform distributions are used to construct the prior (Martin et al., 2011; Pastor-Poquet et al., 2019; Tolessa et al., 2023). This choice is usually motivated by the lack of prior knowledge on the parameter values beyond their plausible range, hence the use of a so called uninformed prior. On the other hand, covariance plays a crucial role in bypassing AD models 'identifiability issues. Gaussian distributions offer a simple way to model covariance while uniform distributions do not. To conciliate flat priors with covariance, new parametrizations of AD models could be considered. For instance, parametrizations considering the ratio of the maximum growth rate and Monod constant might reduce the need for correlations. Another option could be reparameterizations where a the gaussian prior is translated into a uniform prior (using gaussian quantiles transform). Accumulating information about the actual prior distribution of the parameters, as observed, would inform the best practical choice.

*3.9. Applicability to other models*

Although VarBUQ was only evaluated on AD models, it should have similar performance when applied to models involving kindred mechanisms. Most biochemical reaction network models display similar features, relying on a combination of ODEs and algebraic equations, and using similar formulas to infer reaction kinetics from the concentration of reactants. For models focusing on microbial communities (e.g., AM2, ADM1, ASM2, models for dark fermentation, etc.), the network usually corresponds to Monod equations in cascade, with corrections for the impact of environmental parameters such as pH or temperature. For cell-centered models (e.g., dynamic metabolic simulation), the same approach is implemented through e.g. Michaelis-Menten kinetics which are mathematically analogous to the Monod equation. Thus, it could be considered that all these models form a family with comparable non-linearity and differing by their complexity, that is to say the number of represented reactions and model parameters. VarBUQ should display similar advantages and limits for models belonging to this family.

## 4. Conclusions

UQ is crucial to ensure that the right level of confidence is given to future model predictions. The Bayesian-inspired methodology outperformed the most commonly used UQ techniques, both regarding parameter recovery and confidence intervals on test predictions. It benefits from inductive bias encoded in the prior, mitigating the risk of overfitting, and improving robustness compared to standard calibration. Its computational cost, while important, was still sufficiently small to be used in real world scenario and could still be further improved. The methodology is implemented in a readily available python package to facilitate future use.

E-supplementary data for this work can be found in e-version of this paper online.

Data and code availability

The implementation and datasets used are available on the following github repository: https://github.com/BayesianUncertaintyQuantifAnaeroDig/AnaeroDigUQ.

**Table 1.** Assessment of UQ methods for parameters estimation

| Dataset | Bootstrap | FIM | Beale | VarBUQ ($\lambda$) | VarBUQ ($2\lambda$) | VarBUQ ($8\lambda$) |
|---|---|---|---|---|---|---|
| AM2 LN | 0.0 (< 1.2e-2) | 0.0 | 5.0e-9 | **0.91** | n.a | n.a. |
| AM2 HN | 0.0 (< 1.2e-2) | 3.9e-4 | 5.6e-4 | **0.87** | n.a. | n.a. |
| AM2 LF | 0.0 (<1.2e-2) | 1.9e-2 | **0.34** | **0.90** | n.a. | n.a. |
| AM2 HF | 3.1e-2 (< **5.6e-2**) | 2.5e-2 | 3.8e-4 | **0.91** | n.a. | n.a. |
| ADM1 LN | n.a. | 0.0 | 3.3e-11 | 2.5e-6 | 5.2e-4 | **0.56** |
| ADM1 HN | n.a. | 0.0 | 0.0 | **0.22** | **0.48** | **0.89** |
| ADM1 LF | n.a. | 0.0 | 1.1e-3 | 0.0 | 7.1e-9 | 3.7e-3 |
| ADM1 HF | n.a. | 0.0 | 0.0 | 0.0 | 2.5e-8 | 4.1e-3 |

p-values in bold imply that the true set of parameters was inside the 95% confidence region. For the Bootstrap method, the upper bound given in parenthesis is valid with probability at least 0.95. Since datasets LVN (resp. LVF) share its training data and true parameter with dataset LN (resp. LF), the performance of the uncertainty quantification routines is identical.

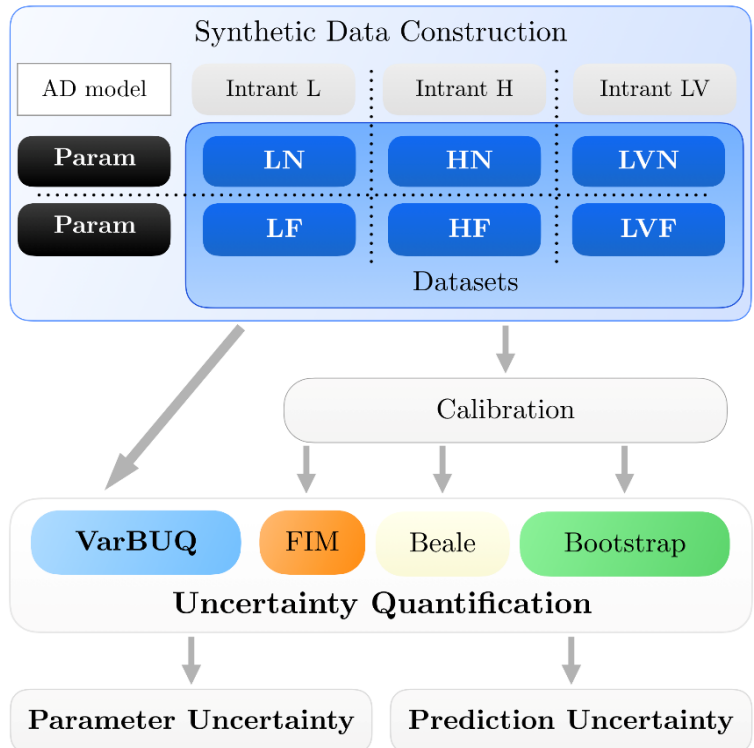

**Figure 1:** UQ analysis methodology. For each AD model, six datasets are evaluated, spanning a choice of two parameters and three intrant descriptions. After calibration, three ad-hoc UQ methods are assessed and compared to VarBUQ's joint calibration and UQ, both on their ability to encompass the true set of parameter values and to quantify uncertainty on the predictions.

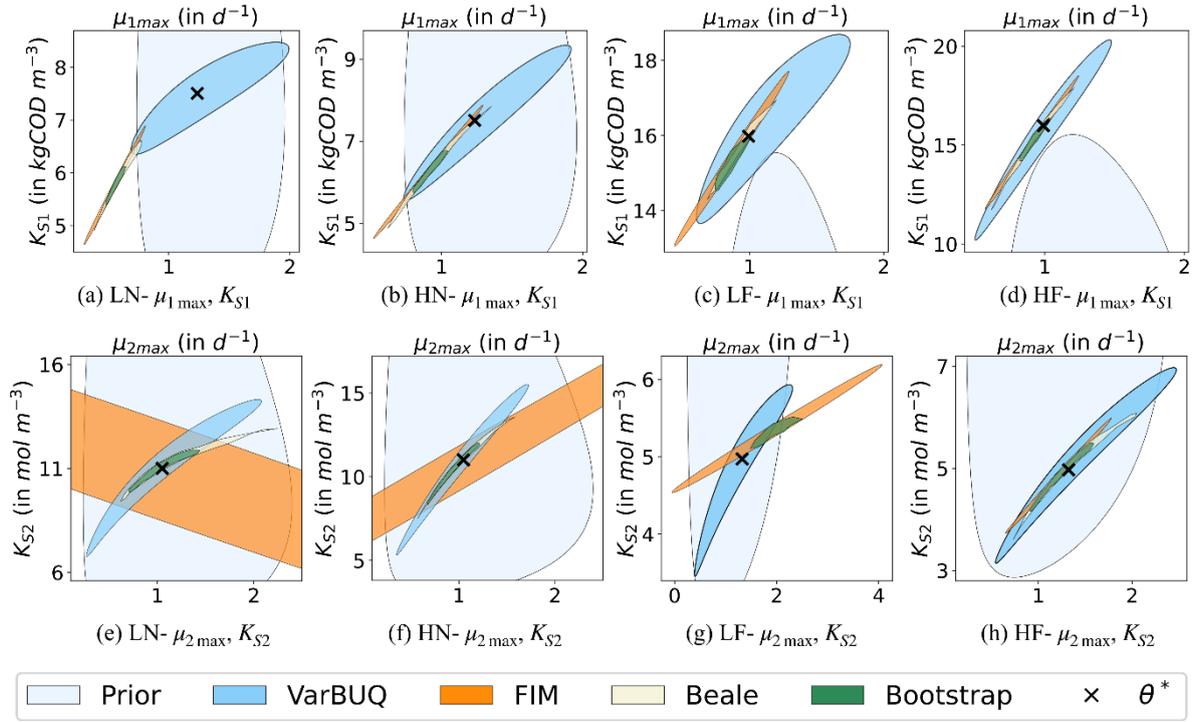

**Figure 2.** 95% Confidence regions for AM2. The prior distribution is in light blue, the posterior in blue, FIM in orange, Beale in beige and Bootstrap in green. The true parameter is represented by a black cross. VarBUQ was the only methodology able to recapture the true sets of parameters in all settings. The methodology can benefit from the prior's inductive bias (figs. 2a, 2b and 2g), but is also able to adapt to cases where the parameter is outside the boundary of the prior's confidence region (figs. 2c and 2d). Those confidence regions are shaped as ellipses in log-space. The ellipsoidal confidence regions obtained through FIM tend to englobe those constructed through Beale's or Bootstrap methods. They suffer from some instability as exhibited in figs. 2e and 2f, englobing negative values. The confidence regions obtained through Beale's method tend to englobe those constructed through the Bootstrap method. Both methods regions with similar curvatures and an overall direction coherent with FIM's confidence regions, except for fig. 2e. All UQ methods responded to the limited identifiability of the maximum growth rate and Monod constant (i.e., the fact that those parameters can compensate for one another) by constructing confidence regions which are squeezed along an axis.

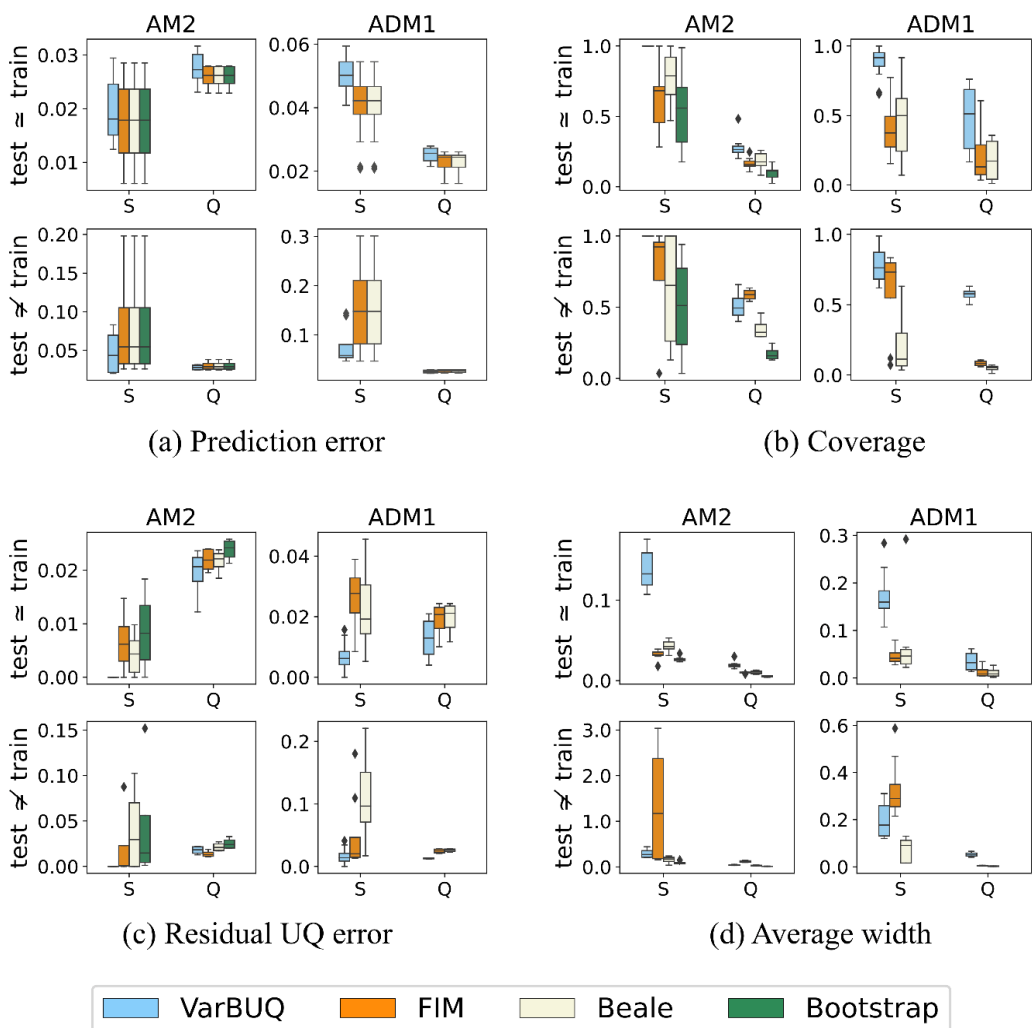

**Figure 3**. Performances of UQ methods on predictions (VarBUQ in blue, FIM in orange, Beale in beige and Bootstrap in green). Types of predictions are grouped depending on whether they are soluble compounds (S) or gas flows (Q). VarBUQ obtained slightly larger test error when the test dataset is similar to the train dataset, but noticeably lower test error when the test dataset exhibit distribution shift (fig. 3a). The coverage of VarBUQ's confidence intervals on the predictions was globally higher than those of the remaining methods - achieving systematically 100% coverage for soluble compounds for AM2 (fig. 3b). The coverage and residual error after projection on the confidence intervals are globally coherent with the width of the confidence intervals, with the notable exception of FIM's behaviour for soluble compounds under distribution shift, where the higher width of the confidence intervals does not result in higher coverage or smaller residual errors compared VarBUQ.

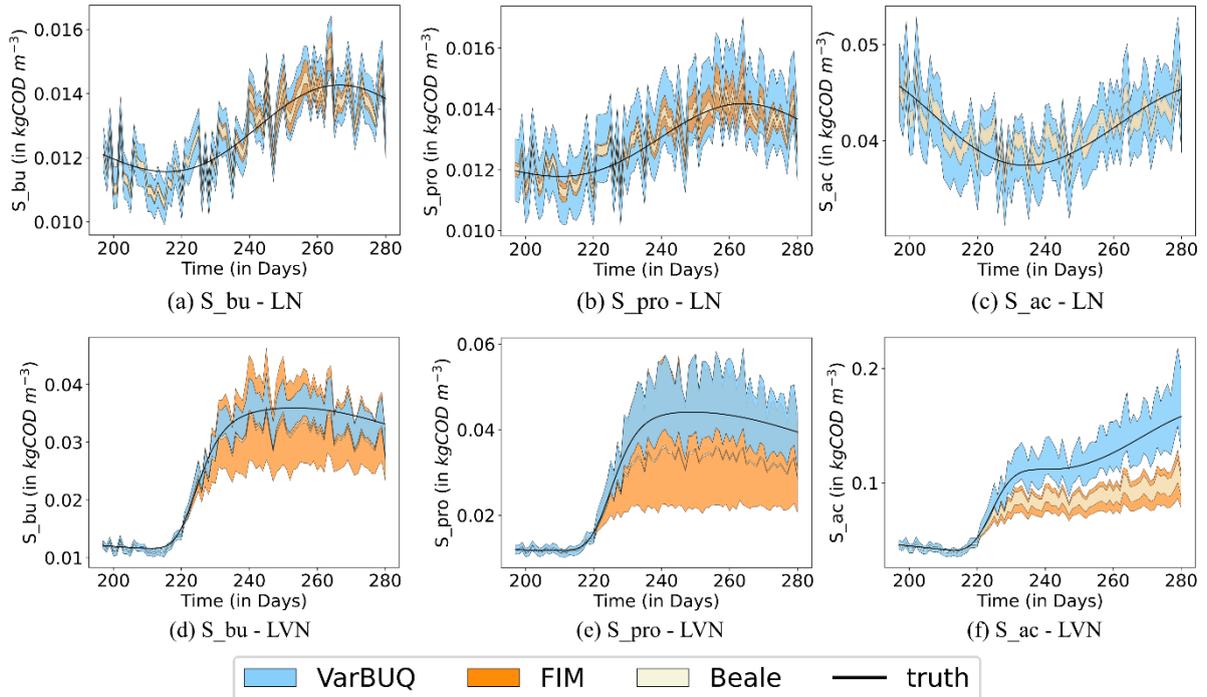

**Figure 4.** 95% Confidence intervals on predictions for the main three VFAs for LN (first row) and LVN (second row) datasets. The confidence intervals from all methods include high frequencies absent from the true signal, due to the noisy intrant description used. When the test set is similar to the training set, the Bayesian calibration has wider confidence intervals, mitigating the impact of the intrant noise (figs. 4a to 4c). Under distribution shift, FIM's confidence intervals widen considerably, englobing the true signal for both butyrate and propionate concentration, while Beale's confidence interval remains centered tightly around the inadequate mean prediction (figs. 4d and 4e). Both FIM and Beale's confidence intervals are unable to account for the increase of acetate production, contrary to VarBUQ's calibration (fig. 4f). Figures for the remaining datasets are available on the github repository https://github.com/BayesianUncertaintyQuantifAnaeroDig/AnaeroDigUQ.